\documentclass[a4paper,fleqn,usenatbib,letter]{mnras}

\usepackage{color}

\usepackage[T1]{fontenc}
\usepackage{ae,aecompl}

\usepackage{amsmath}
\usepackage{grffile}
\usepackage{multirow}
\usepackage[flushleft]{threeparttable}
\usepackage{booktabs}

\def\SFR{{\rm SFR}}
\def\sSFR{{\rm sSFR}}
\def\SFC{{\rm SFC}}

\usepackage{url}
\usepackage{times}

\usepackage{graphicx}

\usepackage{amssymb,amsmath}

\newcommand{\acknowledgements}{\begin{small}\section*{Acknowledgments}\end{small}}

\title[Are star formation rates of galaxies bimodal?]{\vspace{-0.3cm}Are star formation rates of galaxies bimodal?}

\author[R. Feldmann et al.]{
Robert Feldmann$^{1}$\thanks{E-mail: feldmann@physik.uzh.ch}
\\
$^{1}$Institute for Computational Science, University of Zurich, Zurich CH-8057, Switzerland\\
}

\begin{document} 

\maketitle 

\begin{abstract}
Star formation rate (SFR) distributions of galaxies are often assumed to be bimodal with modes corresponding to star-forming and quiescent galaxies, respectively. Both classes of galaxies are typically studied separately and SFR distributions of star-forming galaxies are commonly modelled as lognormals.
Using both observational data and results from numerical simulations, I argue that this division into star-forming and quiescent galaxies is unnecessary from a theoretical point of view and that the SFR distributions of the whole population can be well fitted by zero-inflated negative binomial distributions. This family of distributions has three parameters that determine the average SFR of the galaxies in the sample, the scatter relative to the star-forming sequence, and the fraction of galaxies with zero SFRs, respectively. The proposed distributions naturally account for (i) the discrete nature of star formation, (ii) the presence of `dead' galaxies with \emph{zero} SFRs, and (iii) asymmetric scatter.
Excluding `dead' galaxies, the distribution of $\log{}{\rm SFR}$ is \emph{unimodal} with a peak at the star-forming sequence and an extended tail towards low SFRs. However, uncertainties and biases in the SFR measurements can create the appearance of a bimodal distribution.
\end{abstract}

\begin{keywords}
galaxies: evolution -- galaxies: formation --  galaxies: star formation.\\[-0.7cm]
\end{keywords}

\section{Introduction}
\label{sect:introduction}
The bimodal colour distribution of nearby galaxies leads to a natural classification into `blue' and `red' galaxies (e.g., \citealt{Baldry2004a, Brammer2009}). On average, blue (red) galaxies in the local Universe have high (low) star formation rates (SFRs) although dust extinction complicates this basic picture \citep{Whitaker2012b, Taylor2014}. Hence, galaxies are often divided based on their level of star formation activity into star-forming and quiescent galaxies (e.g., \citealt{Balogh2004a, Moustakas2013}). However, whether the distribution of SFRs is also bimodal remains an open question (e.g., \citealt{Elbaz2007, McGee2010}).

The SFRs of star-forming galaxies strongly correlate with their stellar masses resulting in a well-defined `star-forming sequence' \citep{Noeske2007d}. In contrast, quiescent galaxies have generally very low (or vanishing levels) of SFRs whose exact amount is challenging to infer observationally (e.g., \citealt{Brinchmann2004b, Utomo2014b, Chang2015}). Hence, most observational studies focus on the SFRs of `star-forming galaxies' alone and model their distribution at fixed stellar mass with a lognormal distribution (e.g., \citealt{Noeske2007d}) or with the sum of two lognormal distributions \citep{Sargent2012}. The intrinsic scatter around the star-forming sequence is found to be about $\sim{}0.3-0.4$ dex, essentially independent of redshift ($z\sim{}0-6$; e.g., \citealt{Chang2015, Salmon2015b, Schreiber2015b, Shivaei2015}). 
The approach of approximating the distribution of SFRs of star-forming galaxies with a lognormal is also adopted in most analyses of galaxy formation simulations 
(e.g., \citealt{Schaye2015, Sparre2015a, Dave2016, Feldmann2016}).

However, this approach has a number of shortcomings. First, a clear separation of galaxies into star-forming and quiescent is challenging in practice. Classifications based on colour--magnitude diagrams suffer from a large population of dust-obscured star-forming galaxies with colours intermediate between blue and red \citep{Salim2009, Taylor2014, Chang2015}. 
While colour--colour diagrams offer a more robust alternative (\citealt{Wuyts2007}), 
the mapping from colours to SFRs can be biased by relatively small amounts of recent star formation, by dust, and by the presence of evolved stellar populations (e.g., \citealt{Salim2009, Wuyts2011, Fumagalli2014}). For instance, while high-redshift galaxies classified as quiescent based on colour--colour diagrams have typically significantly reduced SFRs \citep{Man2016, Straatman2016}, some of them show non-negligible levels of SFR and dust extinction (e.g., \citealt{Brammer2009, Spitler2014, Mancini2015b}). It is thus legitimate to ask whether quiescent and star-forming galaxies are actually two separate populations or whether galaxies simply form a continuum from low to high specific SFRs (sSFRs) without a natural dividing point.

Secondly, a lognormal distribution predicts a symmetric scatter around the star-forming sequence, in contrast to the predictions of many observational and simulations studies \citep{Brinchmann2004b, Chang2015, Dave2016}. Often this difference is attributed to an `imperfect separation' into star-forming and quiescent galaxies (e.g., \citealt{Chang2015}). Instead, I argue that the level of asymmetry of the scatter contains important information about galactic star formation and should not be ignored.

Thirdly, star formation is correlated in space and time as stars are typically born in clusters \citep{Lada2003}. These correlations introduce a level of discreteness into the star formation process. For instance, star clusters in the Milky Way have masses ranging from about $\sim{}10^2-10^3$ $M_\odot$ for open clusters to $\sim{}10^5$ $M_\odot$ for young massive clusters \citep{PortegiesZwart2010a}. Star clusters can be even more massive in star-bursting galaxies \citep{Zhang1999} and at high redshift, when galaxies are generally more gas rich (e.g., \citealt{Swinbank2010, Kruijssen2012a}). The discreteness effect is further exacerbated in observationally inferred SFRs as those trace energy injections from rare, massive stars \citep{Kennicutt1998}. A lognormal distribution, which is a continuous distribution, does not account for this discrete mode of star formation. 

Finally, the star formation activity, especially for higher redshift and/or lower mass galaxies, is thought to be bursty (e.g., \citealt{Dominguez2015, Sparre2015}). Consequently, many galaxies may experience intermittent episodes of low or vanishing SFR \citep{Feldmann2017}. However, galaxies with fully suppressed SFRs cannot be modelled by a lognormal distribution. Thus, in practice, galaxies with SFRs below the detection limit are excluded from (non-stacked) analyses of the star-forming sequence even if such galaxies are star forming according to their colours (e.g., \citealt{Whitaker2014b}).

As I argue in this Letter, these shortcomings can be mitigated by dropping the assumption of a lognormal SFR distribution and by not dividing galaxies into star-forming and quiescent galaxies in the first place. In particular, I propose to replace the lognormal ansatz with (zero-inflated) negative binomial distributions (NBDs). This family of distributions found wide applicability in particle physics (e.g., multiplicity distributions of charged particles in hadronic collisions; \citealt{Alner1985}), astrophysics (e.g., the number of globular clusters in galaxies and event rates of fast radio bursts; \citealt{DeSouza2015, Wiel2016}), and cosmology (e.g., modelling count-in-cell distributions and void probability functions; \citealt{Carruthers1983, Gaztanaga1992}) but, to my knowledge, has not been used to model distributions of SFR or sSFRs.

Choosing an appropriate model for the distributions of SFRs is not self-evident as the origin of the scatter around the star-forming sequence is not well understood. As gas accretion on to galaxies and star formation within galaxies are likely linked (e.g. \citealt{Bouche2010, Almeida2014}), the scatter may be related to variations in the gas accretion rates \citep{Dutton2010, Forbes2014c} or halo growth rates \citep{Feldmann2015, Feldmann2016, Rodriguez-Puebla2016}. The scatter may also arise from random stochasticity \citep{Kelson2014b}, gas fraction variations (e.g., \citealt{Magdis2012, Saintonge2012c, Scoville2016}), changes in the efficiency of star formation (e.g., \citealt{Genzel2010, Saintonge2012c}), a natural diversity in star formation histories \citep{Gladders2013, Dressler2016}, or combinations of some of these processes (e.g., \citealt{Tacchella2016, Feldmann2017}). 
Hence, it appears justified to explore empirically how well SFRs  follow various basic distributions. 

With the recent availability of large numbers of reliable SFR measurements these basic distributions can be compared against observations and numerical simulations. I use spectral energy distribution (SED) based SFR estimates of galaxies in the local Universe based on the Sloan Digital Sky Survey (SDSS) \citep{Chang2015}, ultraviolet (UV) and infrared (IR) based SFR measurements of galaxies at $z\sim{}2$ from 3D-HST \citep{Brammer2012b, Skelton2014}, as well as SFRs measured in cosmological simulations \citep{Feldmann2016} that are part of the \emph{Feedback in Realistic Environments} (FIRE) Project \citep{Hopkins2014}.

This Letter is organized as follows: Section \S\ref{sect:probdist} introduces probability distributions to model SFRs in galaxy samples. Section \S\ref{sect:comparison} shows that zero-inflated NBDs (zNBDs) provide adequate approximations to SFR distributions in observations and numerical simulations. I discuss the implications for a possible SFR bimodality in the final section.\\[-0.9cm]

\section{Modelling the distribution of SFRs}
\label{sect:probdist}

\begin{figure*}
\begin{tabular}{ccc}
\includegraphics[height=50mm]{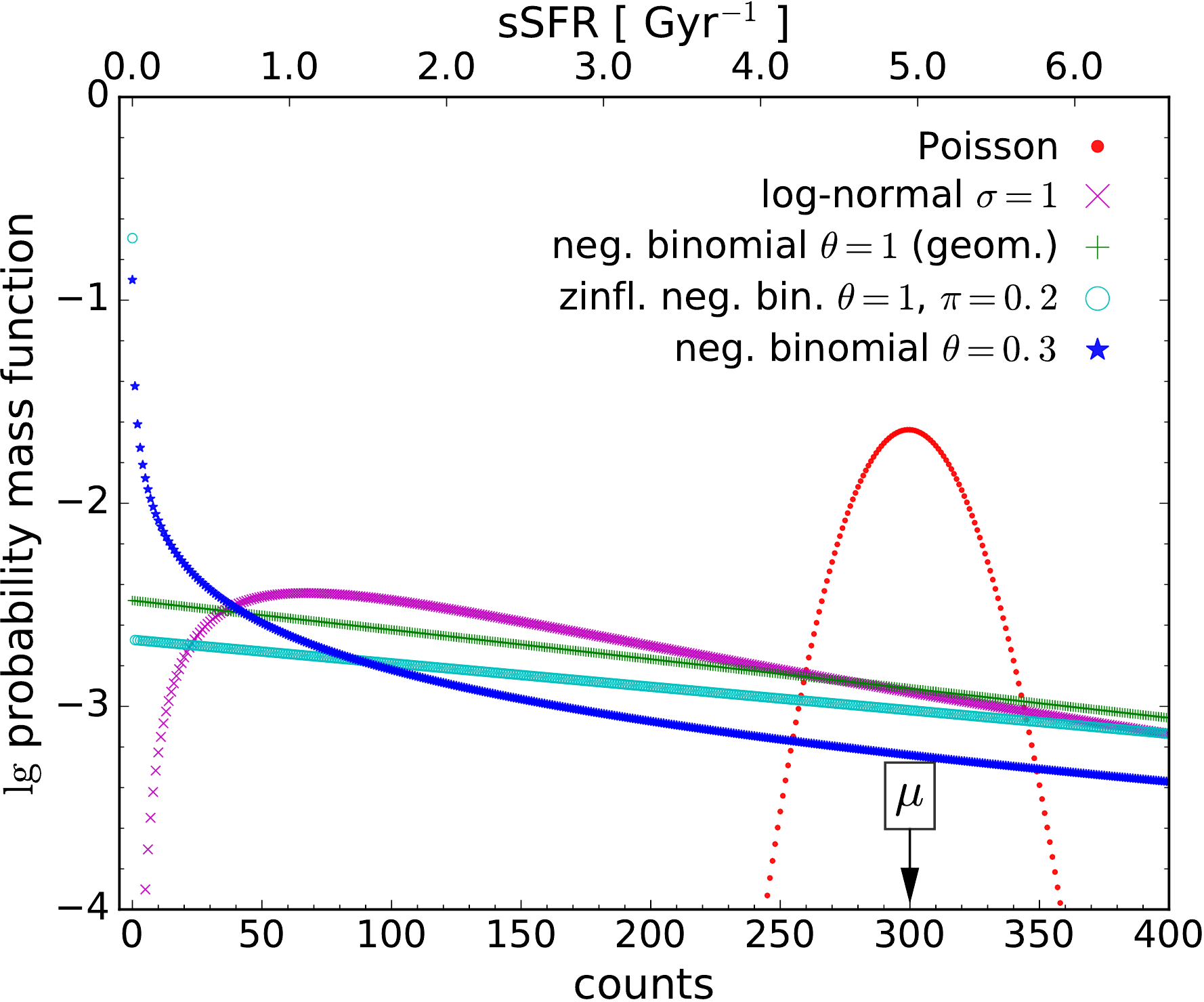} &
\includegraphics[height=50mm]{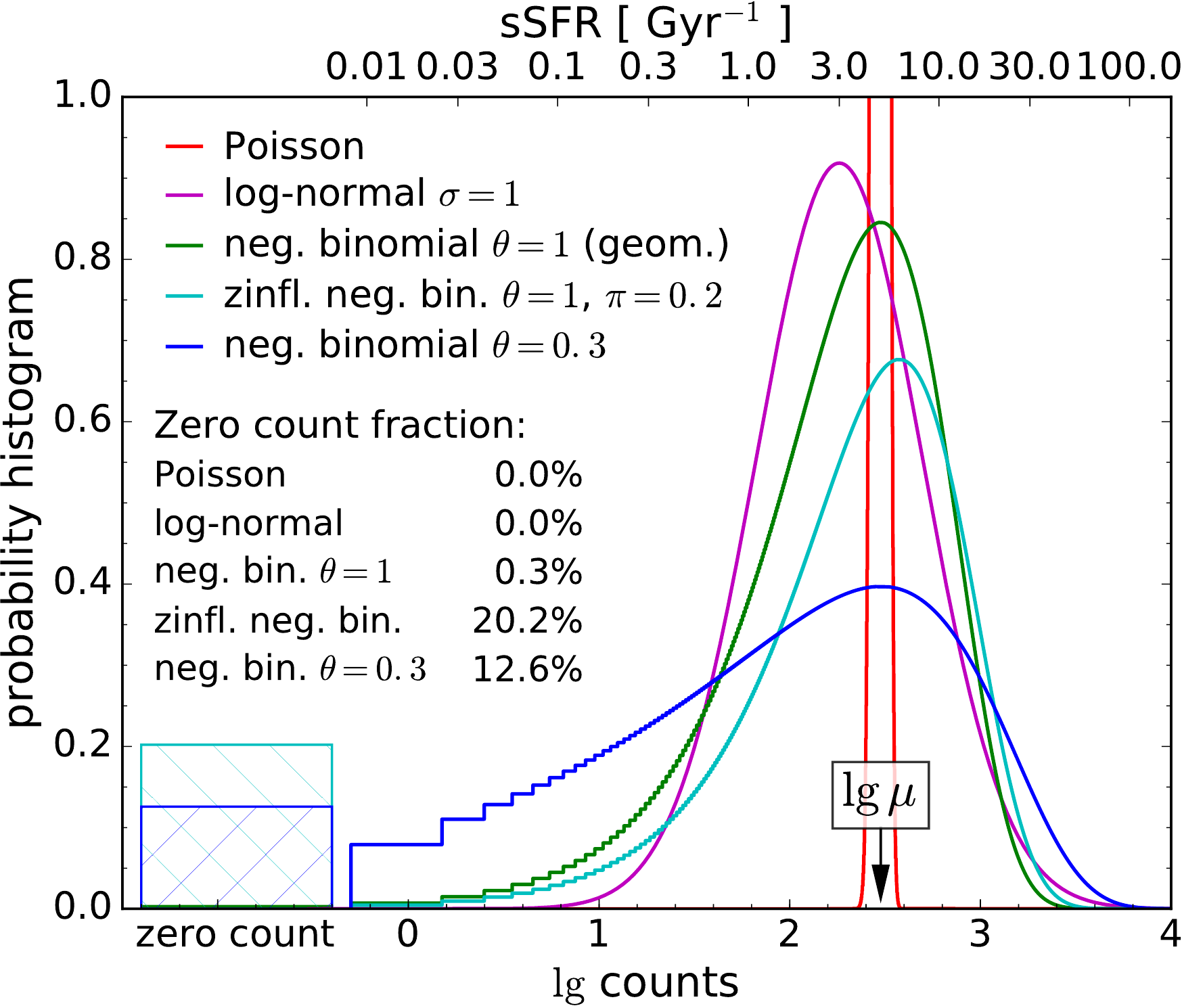} &
\includegraphics[height=45.5mm]{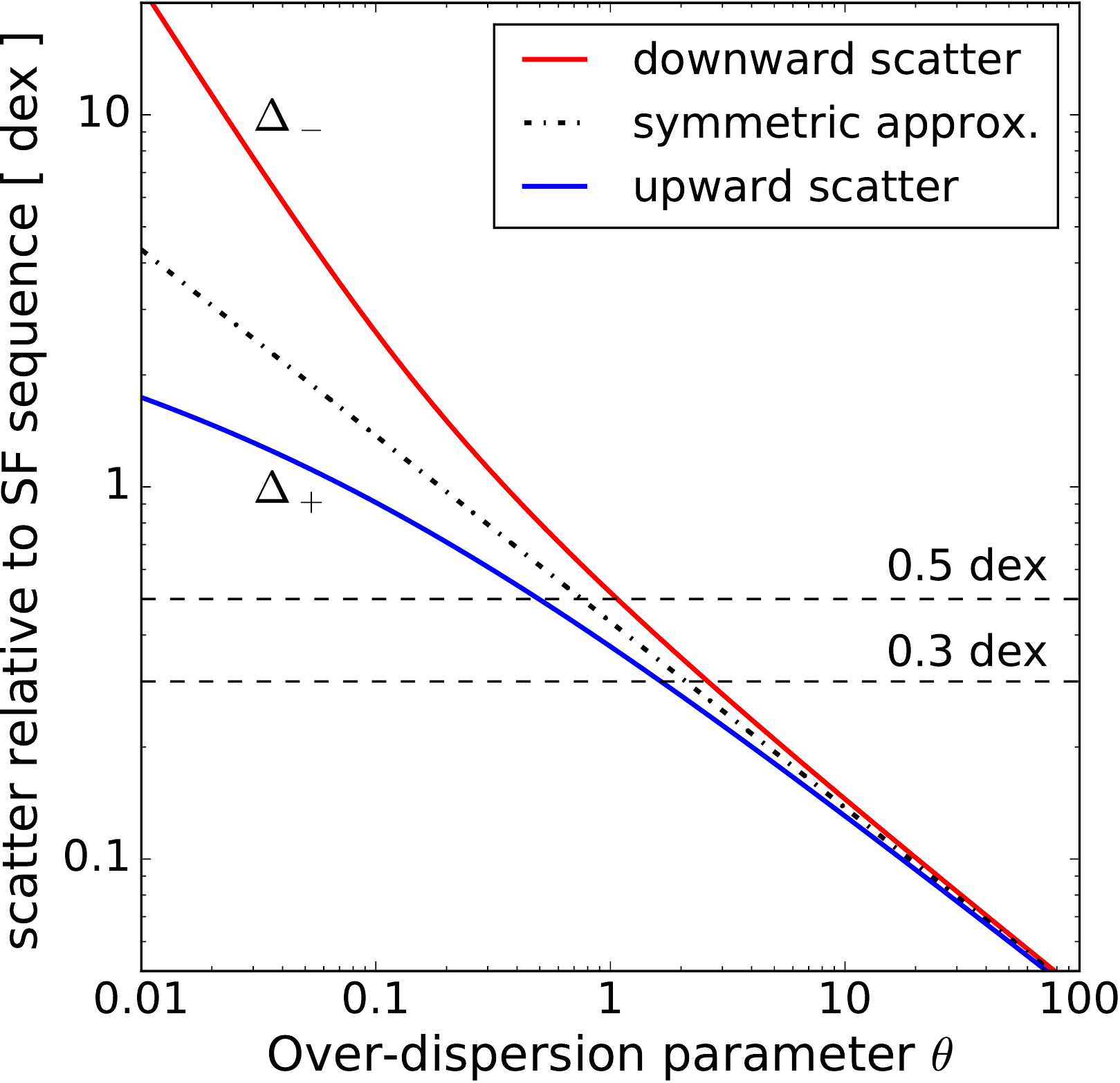} \\
\end{tabular}
\caption{\emph{Properties of negative binomial, zero-inflated negative binomial, lognormal, and Poisson distributions.} All distributions have the same mean number of counts ($\mu=300$). For illustration purposes, I convert between counts (bottom axis) and sSFR (top axis) by assuming that each count corresponds to a fractional increase of galactic stellar mass by $3.3\times{}10^{-4}$ over the past 20 Myr.
Left-hand panel: PMFs of all considered distributions are unimodal. The PMF of an NBD with $\theta\leq{}1$ decreases monotonically with increasing counts. For large $\mu$, the PMF of a Poisson distribution approximates a normal distribution with mean = variance = $\mu$.
Middle panel: probability density function (PDF) of \emph{log counts}. All distributions show a well-defined 'star-forming sequence' with a peak close to $\lg\,\mu$. Compared with the observed star-forming sequence, a Poisson distribution predicts a much narrower star-forming sequence. NBDs with $\theta=1$ predict a reasonable amount of scatter and a negligible fraction of galaxies with zero SFRs. NBDs with $\theta<1$ result in a strongly broadened star-forming sequence and in a significant number of non-star-forming galaxies. Zero-inflated variants of the shown distributions boost the probability of having zero counts.
Right-hand panel: upward ($\Delta_+$) and downward ($\Delta_-$) scatter relative to the peak of the $\lg\,{\rm count}$ distribution as a function of $\theta$ for NBDs (see the text). The downward scatter is generally larger than the upward scatter. The dot--dashed curve shows the approximation $\theta^{-1/2}/\ln{}10$, which holds if $1\lesssim{}\theta\ll{}\mu$. NBDs with $\theta\sim{}0.3 - 3$ have a scatter relative to the peak of the star-forming sequence of about $0.2-1$ dex.
}
\label{fig:Dist}
\end{figure*}

At fixed stellar mass, SFRs of star-forming galaxies are typically assumed to obey lognormal distributions (e.g., \citealt{Chang2015}), i.e., the logarithm\footnote{In the following, $\log$, $\lg$, and $\ln$ denote the logarithm to an arbitrary base, to base 10, and to base ${\rm e}$, respectively.} of the $\SFR$ is assumed to be a continuous variable that is normally distributed with standard deviation $\sigma\equiv\sigma_{\ln\SFR}=\sigma_{\lg\SFR}\ln{}10$. The mean, median, and most probable value of $\lg\SFR$ coincide, and they define the position of the star-forming sequence for the given stellar mass. 

Instead, I propose to model SFRs of galaxies with NBDs. As NBDs describe count data, I assume that the star formation activity over time $t_{\rm av}$ consists of individual star formation events, each adding mass $m_{\rm SFC}$, i.e., 
\begin{equation}
\SFR=\SFC\,m_{\rm SFC}/t_{\rm av}.
\label{eq:SFCtoSFR}
\end{equation}
$\SFC$ (`star formation count') is a non-negative integer-valued random variable with a [potentially zero-inflated (discussed later)] NBD.

The probability mass function (PMF) of an NBD is specified by two parameters, e.g., the expected count $\mu$ and a shape parameter $\theta$, both positive real numbers. The probability of the outcome $\SFC=k\in\mathbb{N}$ is (e.g., \citealt{Cameron2013})
\begin{equation}
P_{\rm NB}(k; \mu, \theta)=\binom{\theta+k-1}{\theta-1}\left(\frac{\mu}{\theta+\mu}\right)^k\left(\frac{\theta}{\theta+\mu}\right)^\theta.
\label{eq:PMF_negbin}
\end{equation}
If $\theta$ is a positive integer, eq. (\ref{eq:PMF_negbin}) describes the probability distribution of the number of Bernoulli trials (each with a success probability $\theta/[\theta+\mu]$) undertaken before the $\theta$th trial is successful. Cases with $\theta=1$ are known as geometric distributions. There exist many other characterizations of NBDs \citep{Boswell1970}. For the remainder of this Letter, I will assume $\theta=O(1)$ and $\mu\gg{}\theta$.

NBDs are often used to model count data that is `overdispersed' relative to a Poisson distribution\footnote{A Poisson-distributed random variable 
has a variance equal to its mean value $\mu$. In contrast, the variance of a random variable with an NBD is equal to $\mu + \mu^2/\theta$, i.e., the variance at given $\mu$ can be adjusted by changing $\theta$.}. NBDs are Poisson--Gamma mixtures which suggests a simple physical interpretation: The overdispersion arises from galaxy-by-galaxy variations of the expected number of SFCs even among galaxies of the same stellar mass. This is plausible as the expected number of SFCs in a galaxy should vary depending on the state of its interstellar medium.

Zero-inflated models \citep{Mullahy1986, Lambert1992} increase the probability of obtaining zero counts, i.e., of producing galaxies with zero SFRs. For zNBDs,
\begin{equation}
P_{\rm zNB}(k; \pi, \mu, \theta) = \pi\,\delta{}_{0k} + (1-\pi)P_{\rm NB}(k; \mu, \theta),
\end{equation}
where $\pi\in{}[0, 1]$ parametrizes the excess probability to obtain zero counts: $P_{\rm zNB}(0; \pi, \mu, \theta) - P_{\rm NB}(0; \mu, \theta) = \pi\,[1-P_{\rm NB}(0; \mu, \theta)]$. Candidate processes responsible for $\pi>0$ include galactic outflows powered by starbursts or active galactic nuclei (e.g. \citealt{King2015, Somerville2015}, and references therein) and strong environmental effects \citep{Gunn1972}.

Examples of lognormal, negative binomial, zero-inflated negative binomial, and Poisson distributions are shown in Fig.~\ref{fig:Dist}. For illustrative purposes I convert between PMFs and probability density distributions (PDFs) by approximating probability point masses with intervals of uniform probability density. As the figure shows, a count of zero is the most probable outcome for random variables with (z)NBDs if $\theta\leq{}1$. 
Interestingly, while the probability of obtaining a certain count value is unimodal and decreases monotonically with the count value for (z)NBDs, the distribution of the \emph{logarithm of the count variable} shows a well-defined peak near\footnote{Provided $\mu\gg{}\theta$, the peak of $\lg$ count is near $\lg(\mu+\theta/2)\sim{}\lg\mu$. In contrast, for a lognormally distributed random variable $X$, the distributions of $\lg{}X$ and $X$ peak at $\lg\mu - \frac{\sigma^2}{2\ln{}10}$ and $\mu{}/{\rm e}^{1.5\sigma^2}$, respectively.
}
the logarithm of $\mu$. In addition, a conversion to $\log$ counts requires that zero and non-zero counts are treated as separate components. As I argue in \S\ref{sect:conclusion}, this split into two components may lead to an apparent `bimodality' of $\log$ SFR distributions. 

Fig.~\ref{fig:Dist} also illustrates that the distribution of $\log$ counts around the peak is \emph{asymmetric} for (z)NBDs. There is a significant tail towards lower values, resulting in a non-negligible probability of a zero count outcome if $\theta<1$. Furthermore, the degree of asymmetry and the width of the $\log$ count distribution increases with decreasing $\theta$ (see the right-hand panel). There, I plot how the upward scatter ($\Delta{}_+$) and the downward ($\Delta{}_-$) scatter\footnote{$\Delta_+$ and $\Delta_-$ are defined as follows. Let $\lg{}c_*$ denote the position where the PDF of $\lg\,{\rm counts}$ reaches its maximum value $p_*$. Increasing (decreasing) the counts by $\Delta_+$($\Delta_-$) dex relative to $c_*$ results in a decrease of the PDF by $\chi={\rm e}^{-1/2}$ relative to $p_*$. The factor ${\rm e}^{-1/2}$ is chosen such that $\Delta_+=\Delta_-=1$ standard deviation for a normal distributed PDF of $\lg\,{\rm counts}$. For $\mu\gg{}\theta$, $\Delta_\pm=\lg(-W_\mp(-\chi^{1/\theta}/{\rm e}))$ where $W_+$ and $W_-$ are the principal and the -1 branch of the Lambert $W$ function, respectively.} relative to the peak of the PDF scale with $\theta$. For $\theta<1$, upward scatter and downward scatter differ significantly from each other, while for $\theta\gtrsim{}1$, $\Delta{}_+\sim{}\Delta{}_-\sim{}\theta^{-1/2}/\ln{}10$.

\section{Comparison with observations and simulations}
\label{sect:comparison}

I use three samples of galaxies at different redshifts, with different stellar masses, and from different sources to test whether SFR distributions at fixed stellar mass can be approximated by zNBDs.

The first sample comprises 38246 nearby galaxies from SDSS. SFR and stellar mass estimates are based on multiwavelength (UV to mid-IR) SED modelling \citep{Chang2015}. I select a mass-complete sample of galaxies\footnote{Excluding galaxies with flag $\neq$ 1, i.e., those without reliable aperture corrections, {\it WISE} photometry, or SED fits.}  with stellar masses in the range of $10^{10}-10^{11}$ $M_\odot$ and with redshifts $z<0.05456$. I set the SFR of a galaxy to zero if the best-fitting sSFR is very low ($<3\times{}10^{-12}$ yr$^{-1}$) and the modelling error is at least twice the spread of the star-forming sequence.

The second sample contains 2317 galaxies at $z\sim{}2$ from the 3D-HST survey (\citealt{Brammer2012b, Skelton2014}; catalogue v4.1). SFRs in the catalogue are based on UV$+24\mu{}$m luminosities (SFRs of non-detected galaxies are set to zero) while stellar masses are derived from SEDs fits. I select all galaxies\footnote{Excluding galaxies with star\_flag = 1, near\_star = 1, or use\_phot $\neq$ 1.} from the five available fields with stellar masses in the range of $10^{10}-10^{11}$ $M_\odot$ and with redshifts $z=1.5-2$. For both observational samples, I convert SFR into SFC with the help of eq. (\ref{eq:SFCtoSFR}) and by adopting a conversion factor\footnote{\label{foot:Gamma}This choice corresponds to, e.g., $m_{\rm SFC}=2\times{}10^4$ $M_\odot$ and $t_{\rm av}=20$ Myr. However, provided $k\gg{}1$ and $\mu\gg{}\theta$,  (z)NBDs are well approximated by (zero-inflated) gamma distributions
and the PDF of $\lg\,\SFR$ for SFR>0 approaches $\ln(10)\,x\,P_{\rm Gamma}(x; \alpha=\theta, \beta=1)$ with $x=\theta\,\SFR/\langle{}\SFR\rangle{}$,
i.e., it is independent of the conversion factor.} of $m_{\rm SFC}/t_{\rm av}=10^{-3} M_\odot / {\rm yr}$.

The last sample combines SFRs and stellar masses from cosmological galaxy simulations. Specifically, I use 1648 $z=6$ galaxies with $M_{\rm star}=10^{7}-10^{9}$  $M_\odot$ from the MassiveFIRE simulation suite \citep{Feldmann2016, Feldmann2017}. Gas and stellar components of the simulated galaxies are represented by gas and star particles with masses $m_{\rm b}=3.3\times{}10^4$ $M_\odot$. Star formation occurs probabilistically based on the local conditions of the interstellar medium. Each individual star formation event results in the formation of a star particle of mass $m_{\rm b}$. I refer the reader to \cite{Hopkins2014} for a background on the simulation methodology. Stellar masses and SFRs are measured within radii of $0.1$$R_{\rm vir}$ of the primary dark matter haloes hosting a given galaxy, excluding satellites. 
SFRs are averaged over the past 20 Myr and converted into SFCs via eq. (\ref{eq:SFCtoSFR}) with $m_{\rm SFC}=m_{\rm b}$ and $t_{\rm av}=20$ Myr.

The data sets contain galaxies of a range of stellar masses to increase the sample size. To combine the SFR distribution of galaxies with different stellar masses, I convert SFRs into SFCs and perform generalized linear regressions of the SFCs as a function of $\ln{}M_{\rm star}$ with a $\log$ link function\footnote{\label{foot:analysis}Statistical analyses are carried out with \texttt{R} (\url{https://www.r-project.org}) using the standard \texttt{glm} function to fit Poisson and geometric distributions, the \texttt{glm.nb} function from the \texttt{MASS} package to fit NBDs, and the \texttt{zeroinfl} function from the \texttt{pscl} package to fit zNBDs. An example regression script is provided at \url{http://www.ics.uzh.ch/~feldmann/resources.html}.}. I thus simultaneously fit for the position of the star-forming sequence as a function of stellar mass and constrain the parameters for the assumed distribution of SFCs at fixed stellar mass.

\begin{figure}
\begin{tabular}{c}
\includegraphics[width=76mm]{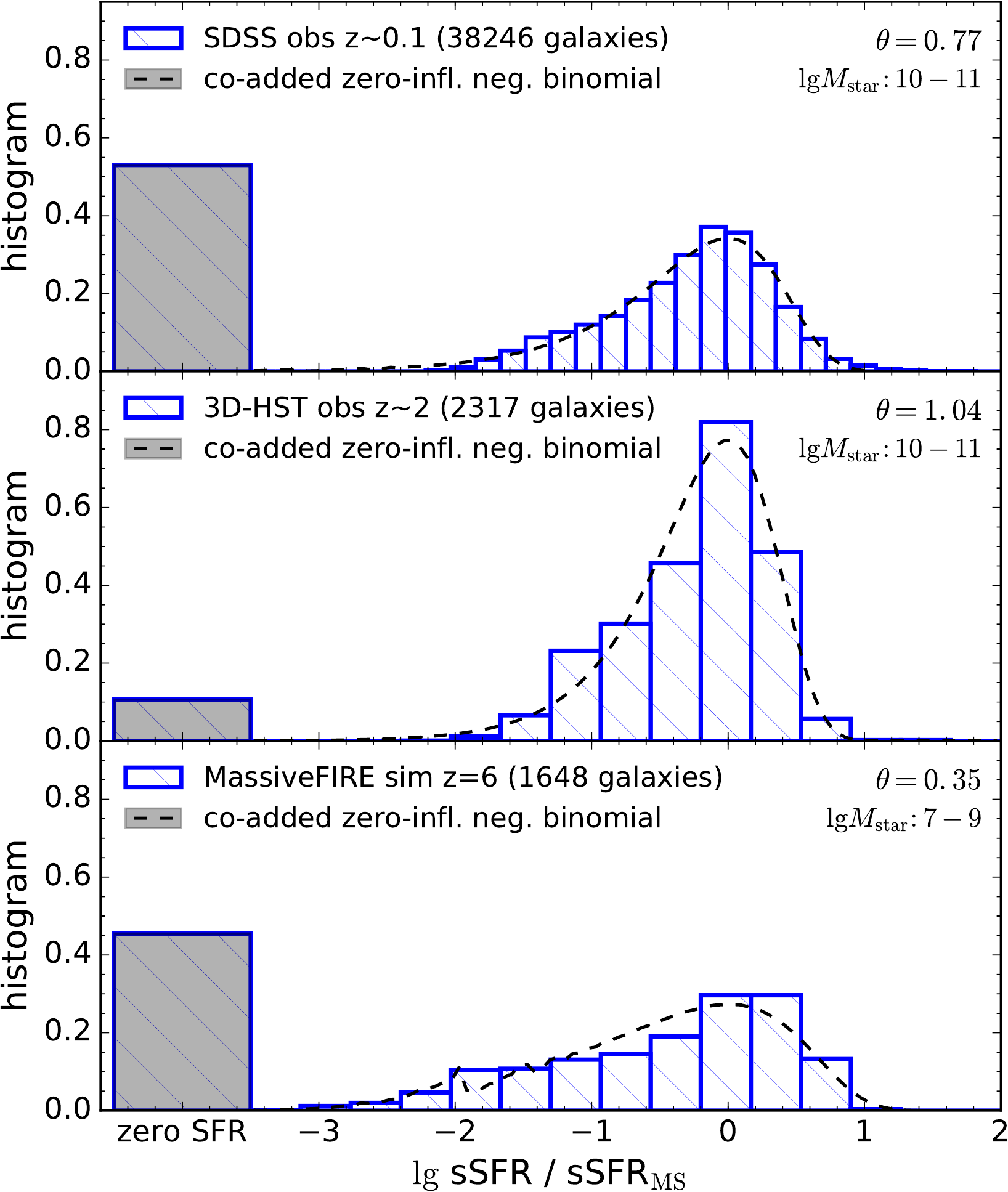}
\end{tabular}
\caption{\emph{Distribution of sSFRs in observations and simulations.} Each panel shows a histogram of the sSFR relative to peak of the star-forming sequence for mass- and redshift-selected samples of galaxies from SDSS \citep{Chang2015}, 3D-HST \citep{Brammer2012b}, and MassiveFIRE \citep{Feldmann2016}.
Best-fitting zNBDs (co-added relative to the $M_{\rm star}$-dependent peak of the star-forming sequence) are shown as dashed lines, and galaxies with undetected or zero SFRs are shown as the grey-shaded areas. The SFR distributions of the three samples are well fitted by zNBDs.}
\label{fig:ObsSim}
\end{figure}

The Akaike information criterion (AIC; \citealt{akaike1974}) measures how well, relative to each other, different statistical models describe a given data set. Among the SFC distributions I tested (negative binomial, geometric, lognormal, Poisson distributions as well as their zero-inflated versions), zNBDs performed best under the AIC metric. 
Fig.~\ref{fig:ObsSim} clarifies why zNBDs work so well. The $\lg\,{\rm sSFR}$ distribution is highly asymmetric with a tail towards low values. Furthermore, there is a significant fraction of galaxies with vanishing SFRs or SFRs below the detection limit. These properties are captured by zNBDs but not by, e.g., lognormal distributions.

I tested the sensitivity of the fitted model parameters (overdispersion parameter $\theta$, stellar mass scalings of the average SFR and of the excess probability $\pi$) by analysing additional MassiveFIRE-based samples for different redshifts and mass resolutions. The main findings are as follows: The slope and normalization of the star-forming sequence show significant changes with redshift (as expected) and only slight changes with particle resolution. The excess probability $\pi$ is not affected by redshift or mass resolution.
The overdispersion $\theta$ does not depend on redshift but varies mildly with mass resolution in a not obviously systematic way.

Finally, I also explored the effect of varying the time interval over which SFRs are averaged. As expected, reducing $t_{\rm av}$ results in a smaller number of SFCs and a larger fraction of galaxies with zero SFRs. Specifically, $\pi$ increases from 20\% to 52\% as $t_{\rm av}$ is reduced from 100 Myr to 5 Myr. Moreover, $\theta$ increases, i.e., the scatter of the star-forming sequence decreases, with increasing averaging time. However, the change in $\theta$ is relatively modest (a factor of 1.9 when $t_{\rm av}$ increases from 20 to 100 Myr). Furthermore, $\theta$ remains unchanged if $t_{\rm av}$ is lowered to 5 Myr. This suggests that the star formation activity of MassiveFIRE galaxies is strongly correlated on $\lesssim{}20$ Myr time-scales.\\[-0.9 cm]

\begin{figure}
\begin{tabular}{c}
\includegraphics[width=80mm]{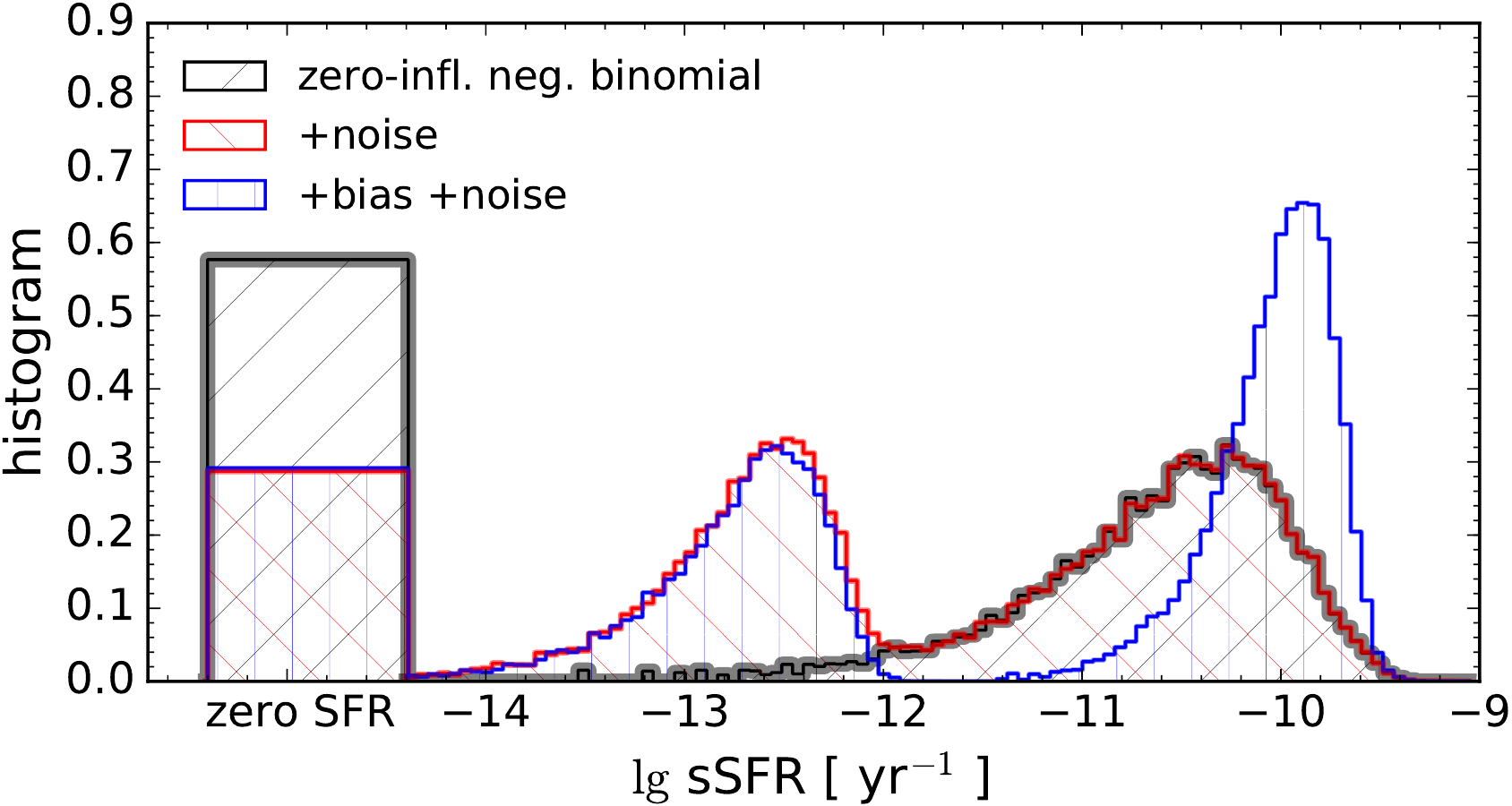}
\end{tabular}
\caption{\emph{Difference between actual SFR distributions and inferred ones.} An input SFR distribution (black histogram) is subjected to measurement uncertainties (red histogram) and non-linear biases (blue histogram). Measurement uncertainties can create the illusion of a bimodal distribution while biases can shift and tighten the appearance of the star-forming sequence and reduce the number of galaxies with intermediate-to-low SFRs.}
\label{fig:Bimodality}
\end{figure}

\section{Discussion and Conclusions}
\label{sect:conclusion}
(Zero-inflated) NBDs predict significant numbers of galaxies with vanishing SFRs. As discussed in \S\ref{sect:probdist}, these `dead' galaxies form a separate component upon log-transforming SFRs, while the remaining galaxies have a unimodal log SFR distribution with an extensive tail towards low SFRs. The reader may ask whether these findings are consistent with claims of an SFR bimodality (e.g., \citealt{Elbaz2007, McGee2010}).

I address this question in Fig.~\ref{fig:Bimodality} in a schematic way. A detailed analysis is left for future work. Using the best fitting zNBD of the SDSS-based sample shown in the top panel of Fig.~\ref{fig:ObsSim}, I create a mock sample consisting of the SFRs of 40000 galaxies with $M_{\rm star}=10^{10.5}$ $M_\odot$ (black histogram). I then subject this sample to non-linear biases and measurement uncertainties\footnote{For the bias, I assume that sSFR below $\sSFR_0=3\times{}10^{-10}$ yr$^{-1}$  are overestimated: $\sSFR_{\rm obs} = \sSFR_0\,(\sSFR/\sSFR_0)^{0.463}$, following \cite{Utomo2014b}. To mimic measurement uncertainties, each SFR is replaced by $\max(0, \SFR + X)$, where $X$ is a normally distributed random variable with zero mean and standard deviation of $0.01$ $M_\odot$ yr$^{-1}$.}. The latter moves galaxies out of the `dead' pile, thereby introducing in a second peak in the $\log$ SFR distribution (red histogram), while biases distort the apparent shape of the star-forming sequence (blue histogram). The combination of measurement noise and bias can create the illusion of a bimodal SFR distribution\footnote{ Similarly, \emph{colour} bimodalities can arise if sSFRs are non-linearly mapped to colours, e.g., if ${\rm colour} = \tanh(3\times{}\sSFR/\sSFR_{\rm MS}-1)$.} with low numbers of galaxies at intermediate-to-low SFRs. 

Fitting the distribution of SFRs with (z)NBDs is straightforward (see footnote \ref{foot:analysis}), and it offers substantial benefits compared with the current standard approach of fitting only the star-forming sub-sample with lognormals. I recommend its use for the modelling of SFR distributions both in observations [e.g., in the simplified form of (zero-inflated) gamma distributions, see footnote \ref{foot:Gamma}] and in simulations (where the conversion parameters $m_{\rm SFC}$ and $t_{\rm av}$ are given).\\[-1.1cm]

\acknowledgements
RF thanks the referee for valuable comments. RF acknowledges financial support from the Swiss National Science Foundation (grant no 157591). 3D-HST observations were taken by the 3D-HST Treasury Program (GO 12177 and 12328) under NASA contract NAS5-26555. Simulations were run under allocations SMD-14-5492, SMD-14-5189, SMD-15-5950 (NASA HEC) and AST120025, AST150045 (NSF XSEDE).  \\[0cm]

\end{document}